\documentclass[useAMS,usenatbib]{mn2e}
\usepackage{subfigure}
\usepackage{graphicx}

\title[Resolving the nuclear dust distribution of NGC 3081]{Resolving the nuclear dust distribution of the Seyfert 2 galaxy NGC 3081}
\author[C. Ramos Almeida et al.]
{\parbox{\textwidth}{C. Ramos Almeida$^{1}$\thanks{E-mail:C.Ramos@sheffield.ac.uk},
M. S\'anchez-Portal$^{2}$,
A. M. P\'erez-Garc\'ia$^{3,4}$,
J. A. Acosta-Pulido$^{3,4}$,
M. Castillo$^{2}$,
A. Asensio Ramos$^{3,4}$,
J. I. Gonz\'alez-Serrano$^{5}$,
A. Alonso-Herrero$^{6}$,
J. M. Rodr\' iguez Espinosa$^{3,4}$,
E. Hatziminaoglou$^{7}$,
D. Coia$^{2}$,
I. Valtchanov$^{2}$,
M. Povi\'c$^{8}$,
P. Esquej$^{6}$,
C. Packham$^{9}$,
B. Altieri$^{2}$
}\vspace{0.4cm}\\
\parbox{\textwidth}{$^{1}$Department of Physics and Astronomy, University of Sheffield, Sheffield, S3 7RH, UK\\
$^{2}$Herschel Science Centre, INSA/ESAC, Madrid, Spain \\
$^{3}$Instituto de Astrof\'\i sica de Canarias, C/V\'\i a L\'{a}ctea, s/n, E-38205, La Laguna, Tenerife, Spain\\
$^{4}$Departamento de Astrof\' isica, Universidad de La Laguna, E-38205, La Laguna, Tenerife, Spain\\
$^{5}$Instituto de F\' isica de Cantabria, CSIC-Universidad de Cantabria, E-39005, Santander, Spain\\
$^{6}$Centro de Astrobiolog\' ia, INTA-CSIC, E-28850, Madrid, Spain\\
$^{7}$European Southern Observatory, Karl-Schwarzschild-Str. 2, 85748 Garching bei M\"unchen, Germany\\
$^{8}$Instituto de Astrof\' isica de Andaluc\' ia (CSIC), Apdo. 3004, 18080, Granada, Spain\\
$^{9}$Astronomy Department, University of Florida, 211 Bryant Space Science Center, P.O. Box 112055, Gainesville, Florida, USA\\}
}

\begin{document}

\date{}

\pagerange{\pageref{firstpage}--\pageref{lastpage}} \pubyear{2011}

\maketitle

\label{firstpage}

\begin{abstract}
We report far-infrared (FIR) imaging of the Seyfert 2
galaxy NGC 3081 in the range 70-500 \micron, obtained with an unprecedented angular resolution,
using the Herschel Space Observatory instruments PACS and SPIRE.
The 11 kpc ($\sim$70\arcsec) diameter star-forming ring of the galaxy appears resolved up to 250 \micron. 
We extracted infrared (1.6-500 \micron) nuclear fluxes, that is active nucleus-dominated fluxes, and  
fitted them with clumpy torus models, which successfully reproduce the FIR emission with small 
torus sizes. 
Adding the FIR data to the near- and 
mid-infrared spectral energy distribution (SED) results in a 
torus radial extent of R$_o$=4$\pm^2_1$ pc, as well as in a flat
radial distribution of the clouds (i.e. the $q$ parameter).
At wavelengths beyond 200 \micron, cold dust emission at T=28$\pm$1 K from the circumnuclear 
star-forming ring of 2.3 kpc ($\sim$15\arcsec) in diameter starts making a contribution to the nuclear emission. 
The dust in the outer parts of the galaxy is heated by the interstellar radiation field (19$\pm$3 K).
\end{abstract}


\section{Introduction}
\label{intro}

The infrared (IR) SED of active galactic nuclei (AGN) 
serves as a sensitive probe of both the dust and the sources that are heating it. 
Dust grains absorb optical and ultraviolet photons from the AGN and from stars and re-radiate them
in the infrared (IR) range.
The IRAS and ISO satellites revealed that Seyfert galaxies are strong FIR and mid-infrared (MIR) 
emitters \citep{Rodriguez87,Spinoglio95} and that this emission is thermal, 
and a combination of a warm, a cold, and a very cold dust components \citep{Radovich99,Perez01}. The warm component is 
produced by dust heated by either the AGN or circumnuclear starbursts (dust at 120-170 K), the cold dust is heated by stars 
in the galaxy disk (30-70 K), and the very cold dust is heated by the general interstellar radiation field (15-25 K). 
With the advent of the Herschel Space Observatory\footnote{Herschel is an ESA space observatory with science instruments provided by 
European-led Principal Investigator consortia and with important participation from NASA.} \citep{Pilbratt10}
it is now possible to map the FIR emission of nearby Seyferts at higher angular resolutions than those previously 
achieved with the Spitzer Space Telescope between 70 and 160 \micron. This, together with the 
unprecedented sensitivities that Herschel offers up to 500 \micron~allow to probe their dust distributions 
at different temperatures. 
In this letter we present new Herschel imaging data of the galaxy NGC 3081, 
which is part of a guaranteed program of FIR imaging observations of Seyfert galaxies. The main goal is 
to characterise their IR SEDs, by 
determining the fractional contributions of the warm, cold, and very cold dust components.   \\
The galaxy NGC 3081 harbours a Seyfert 2 nucleus, although \citet{Moran00} reported a spectacular Type-1 optical spectrum 
in polarised light. The galaxy is at a distance of 32.5 Mpc \citep{Buta98}, which corresponds to 
a spatial scale of 158 pc~arcsec$^{-1}$. 
This early-type barred spiral is forming stars in a series of nested ringlike features: 
%In fact, the galaxy has all four of the main types of resonance rings predicted by models of barred spirals: 
a nuclear 2.3 kpc diameter 
ring (hereafter r1), an inner ring of 11 kpc (r2), an outer ring of 26.9 kpc, and a pseudoring of 33.1 kpc diameter \citep{Buta98,Buta04,Byrd06}.
%In Figure \ref{rings} we show a scheme of the size and orientation of the two inner rings. 

%\begin{figure}
%\centering
%\includegraphics[width=5cm]{buta_mod.ps}
%\caption{Scheme of the size and orientation of the two inner rings of NGC 3081. The r1 ring
%is detected, together with the bar, in the HST images (e.g. Figure 2 in \citealt{Byrd06}). The r2 ring is resolved 
%in the FIR images up to 250 \micron. Figure addapted from \citet{Buta04}.}
%\label{rings}
%\end{figure}

\vspace{-0.6cm}

\section{Observations}

FIR maps of NGC 3081 were obtained with the PACS and SPIRE instruments of the Herschel Space Observatory.
%See \citet{Poglitsch10} and \citet{Griffin10}  for details of the instrument capabilities, calibration methods and
%accuracy of the two instruments. 
The data are part of the guaranteed time proposal ``Herschel imaging photometry of
nearby Seyfert galaxies: testing the coexistence of AGN and starburst activity and the nature of the dusty torus''
(PI: M. S\' anchez-Portal).

The PACS observations were carried out using the ``mini-map'' mode, consisting of two concatenated 3\arcmin~scan line 
maps, at  70$^{\circ}$ and 110$^{\circ}$ (in array coordinates). This 
%at a speed of 20\,arcsec/sec, each one with 10 lines 
%of 3\,arcmin~length and cross-scan step of 4\arcsec. 
%This mode produces a highly homogeneous exposure map within the  central 1\arcmin~area. 
results in a map with a highly homogeneous exposure within the central 1\arcmin~area.
%The set of maps were duplicated to observe through both the 70 \micron~(``blue'') and 
%100 \micron~(``green'') filters. Therefore the galaxy was observed twice through the 160 \micron~(``red'') filter. 
The PACS beams at 70, 100, and 160 \micron~are 5.6\arcsec, 6.8\arcsec, and 11.3\arcsec~full-width half maximum (FWHM) 
respectively. With the SPIRE photometer, the three available bands were observed simultaneously 
using the ``small map'' mode, 
%with two 1\,$\times$\,1 nearly orthogonal scan lines, at a scan speed of 30\arcsec/sec. The scan line length is 
%11.3\arcmin~and the area for scientific use is around 5\arcmin$\times$5\arcmin.
whose area for scientific use is around 5\arcmin$\times$5\arcmin.
The FWHM beam sizes at 250, 350, and 500 \micron~are 18.1\arcsec, 25.2\arcsec, and 36.9\arcsec~respectively.

We carried out the data reduction with the Herschel Interactive Processing Environment (HIPE) v6.0.1951. 
For the PACS instrument, we deemed the extended source version of the standard \textit{PhotProject} reduction script 
as adequate, given the small angular size of the galaxy.  
%This procedure implements a high-pass 
%filtering algorithm to remove the \textit{1/f} noise of the bolometer signals. The sources are masked in 
%order to prevent the high-pass filter to remove flux from extended structures.  
We used the FM v5 photometer response calibration files \citep{Muller11}. 
%The level~2 maps produced by the reduction script 
% for the two scan directions were merged using the \textit{mosaic} task. 
For SPIRE, we applied the standard small 
map script with the `na\"ive' scan mapper task, using the calibration database v6.1. Colour 
corrections (for PACS, see \citealt{Poglitsch10}; please refer to the \citealt{spire11} for 
the SPIRE ones) are small for blackbodies at the expected 
temperatures (e.g. \citealt{Perez01}) and 
have been neglected. More details on the observations and data processing are given in S\'anchez-Portal 
et al. (in preparation).

\begin{figure*}
\centering
\includegraphics[width=15cm]{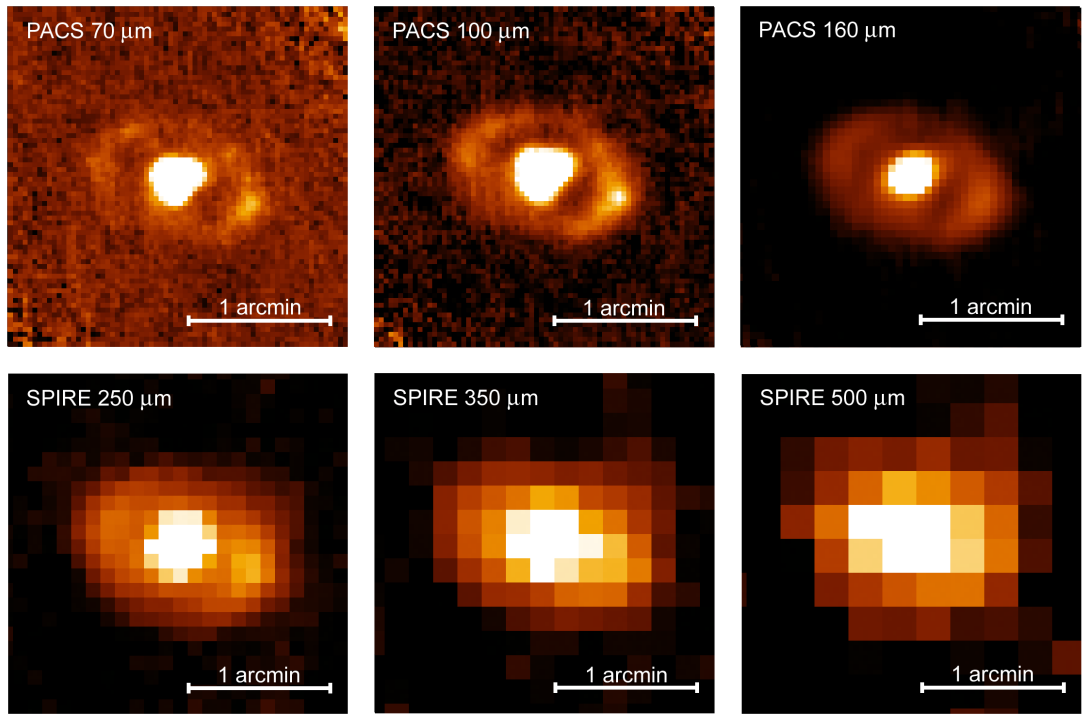}
\caption{Herschel PACS 70, 100, and 160 \micron~images (top) and
SPIRE 250, 350, and 500 \micron~maps (bottom). North is up and East is to the left. 
The r2 ring is resolved up to 250 \micron.}
\label{pacs_spire}
\end{figure*}

The FIR maps of NGC 3081 are shown in Figure \ref{pacs_spire}. The star-forming ring r2 is clearly resolved
in the three PACS images, as well as in the 250 \micron~SPIRE map. There is a brighter region in the 
western side of the ring detected at 70, 100, 160, and 250 \micron~that does not have optical/near-infrared (NIR)
counterpart in the HST images \citep{Buta98,Buta04,Byrd06}. 

To study the IR nuclear emission of NGC 3081, we obtained and compiled unresolved fluxes, 
i.e. either the emission of a point spread function (PSF) component fitted to the data, 
or the emission contained in an aperture diameter equals to the FWHM of the PSF in each band. 
In the case of the FIR, we used GALFIT 2D fitting \citep{Peng02} to obtain the unresolved fluxes. 
For the PACS images we fitted a PSF component, which we identified with the nuclear flux, 
a Sersic profile of R$_e \sim 7\arcsec-18\arcsec$, and a fainter and larger Sersic component of R$_e \sim 45\arcsec-60\arcsec$, 
where R$_e$ is the half-light radius given by GALFIT. For the SPIRE data we only fitted a PSF component
and a Sersic profile of R$_e \sim 75\arcsec-90\arcsec$.  
We also tried this simpler PSF + Sersic fit with the PACS images, but the three-component model results 
in smaller residuals and values of the reduced $\chi^2$ (0.002 for PACS and 0.02 for SPIRE). All the fitted 
Sersic components have indices between 
0.4 and 0.8 (i.e. disk-like) and do not reproduce the r2 ring, which is a residual of the fits.
The PSF input functions are the empirical ones from PACS and SPIRE \citep{Lutz10,Sibthorpe11}.

In addition, we compiled the highest angular resolution NIR and MIR data from the literature to 
construct the nuclear SED. 
Subarcsecond resolution MIR images of NGC 3081 (0.30\arcsec~at 8.74 \micron~and 0.56\arcsec~at 18.3 \micron)
were obtained using the camera/spectrograph T-ReCS on the Gemini-South Telescope. 
The unresolved T-ReCS fluxes from \citet{Ramos09} are reported in Table \ref{psf}, together with  
an additional nuclear flux at 13.04 \micron~from VISIR on the VLT with 
similar resolution as the T-ReCS data \citep{Gandhi09}.
In the NIR, we use the nuclear flux obtained from the NICMOS 1.6 \micron~image 
reported in \citet{Quillen01}. 
%The galaxy was observed using the F160W 
%filter on camera 2 of the NICMOS with a pixel size of 0.075 arcsec and resolution $\sim$0.3\arcsec.

In Table \ref{psf} we report the unresolved NIR, MIR, and FIR fluxes, their uncertainties, and 
the angular resolution at each wavelength. The FIR errors are the result of adding 
quadratically the photometric accuracies (5\% for PACS at 70 and 100 \micron~and 
10\% at 160 \micron; \citealt{Muller10} and 7\% for SPIRE; \citealt{spire11}) 
and the PSF flux determination uncertainties. The latter are the dominant source of error 
and account for the variations in the PSF fluxes associated to the GALFIT fitting in each band.

\begin{table}
%\caption{PACS \& SPIRE perture photometry of NGC 3081}
\begin{tabular}{lccccc}
\hline
Instrument & $\lambda_c$ & \multicolumn{2}{c}{PSF FWHM} & Flux  & Uncertainty \\
 & (\micron) & (arcsec) & (kpc) & (mJy) & (\%) \\
\hline
\hline
NICMOS & 1.6  & 0.20  & 0.032 & 0.22 & 6   \\   
T-ReCS & 8.74 & 0.30  & 0.047 & 83   & 15  \\   
VISIR  & 13.0 & 0.35  & 0.055 & 138  & 10  \\
T-ReCS & 18.3 & 0.56  & 0.083 & 231  & 25  \\	
PACS   & 70   & 5.6   & 0.83  & 758  & 30  \\
PACS   & 100  & 6.8   & 1.01  & 575  & 30  \\
PACS   & 160  & 11.3  & 1.7   & 60   & 30  \\
SPIRE  & 250  & 18.1  & 2.7   & 488  & 50  \\
SPIRE  & 350  & 25.2  & 3.7   & 86   & 50  \\
SPIRE  & 500  & 36.9  & 5.4   & 19   & 60  \\
\hline
\end{tabular}
\caption{IR nuclear fluxes employed in the fit of NGC 3081 with clumpy torus models.
Errors have been obtained by adding quadratically the photometric and PSF subtraction
uncertainties.}
\label{psf}
\end{table}

\vspace{-0.6cm}

\section{Nuclear SED modelling}
\label{modelling}

In a recent series of papers (\citealt{Ramos09}; 2011; \citealt{Alonso11}) 
we fitted the nuclear NIR and MIR emission of Seyfert galaxies with the clumpy torus models 
of \citet{Nenkova08} and were able to nicely constrain the torus parameters, including 
the torus radial extent, $Y = R_o/R_d$, where R$_o$ and R$_d$ are the outer and inner radius of the toroidal distribution 
of clumps respectively. The inner radius is defined by the dust sublimation temperature (T$_d \sim 1500$ K). 
However, it is not clear if the lack of FIR high angular resolution data, which probes 
cooler dust, might bias the fits to smaller torus sizes. In the context of the Nenkova models, 
clouds are heated by the AGN radiation (directly illuminated) and by other clouds (indirectly illuminated)
and each clump contains a range of temperatures itself. The temperature of the dust within the torus scales 
with the square root of the distance to the sublimation radius. Thus, a distribution of clumps with $Y \sim 100$ will 
include dust at temperatures ranging from $T_d$ to $\sim$150 K for the directly illuminated clouds, 
and down to a few Kelvin for the shadowed and more distant ones. 
To test how the addition of FIR data affects the fits described above, 
here we use Herschel PACS and SPIRE nuclear fluxes combined 
with NIR and MIR data of the galaxy NGC 3081 and fitted them with clumpy torus models.
In general, clumpy torus models appear to reproduce better the IR emission of nearby AGN than smooth 
torus models (e.g. \citealt{Alonso03,Mullaney11}) although there is not a general consensus 
on the dust distribution yet. Indeed, in forthcoming publications based on 
guaranteed time Herschel observations of AGN we plan to use the smooth torus models described in 
\citet{Fritz06} and \citet{Hatziminaoglou08} to compare with the results obtained with clumpy torus models.

\begin{figure*}
%\centering
\includegraphics[width=12cm]{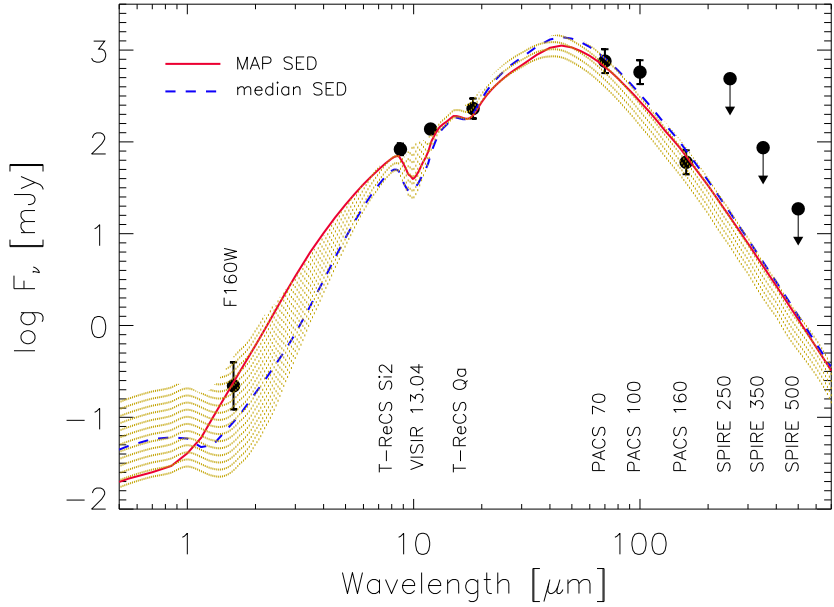}
\caption{Rest-frame IR SED of NGC 3081 (dots).
Solid and dashed lines are the ``best fit'' to the data (MAP)
and the model described by the median of the posteriors respectively.
The shaded region indicates the range of models compatible with the observations at the 2$\sigma$ level.
The SPIRE fluxes are set as upper limits in the fit.}
\label{ngc3081_fit}
\end{figure*}

We constructed the nuclear 1.6-500 \micron~SED of NGC 3081 (see Figure \ref{ngc3081_fit}) using the nuclear 
fluxes reported in Table \ref{psf}. The unresolved NIR and MIR components correspond to a physical 
region of $<$85 pc in diameter and thus, are likely be dominated by emission from the AGN dusty torus. 
The nuclear FIR fluxes of NGC 3081, on the other hand, come from regions with sizes of $\sim$1 kpc in the case 
of PACS and between 2.7 and 5.4 kpc for SPIRE. These regions are much larger than the 
physical scales responsible for the NIR and MIR unresolved emission. 
Indeed, the SPIRE nuclear fluxes include emission from the inner ring of 2.3 kpc diameter (r1), and consequently, we consider them as 
upper limits in the fit (see Figure \ref{ngc3081_fit}). The PACS fluxes exclude r1, but may
include other sources of nuclear emission apart from the torus. This contamination might affect the SED shape  
and consequently, the resulting torus parameters. However, here we work under the assumption that the torus 
is the dominant source of unresolved emission up to 160 \micron. Although \citet{Deo09} showed that, in general, 
the AGN continuum of Seyfert 2 galaxies drops rapidly beyond 20 \micron, NGC 3081 is one of the galaxies in their
sample with the smallest starburst to AGN ratios at 30 \micron. Indeed, the Spitzer IRS flux measurement that they
reported for this galaxy, 1.09 Jy, nicely matches our fitted torus models (see below).

\begin{figure}
\centering
\par{
\includegraphics[width=4cm]{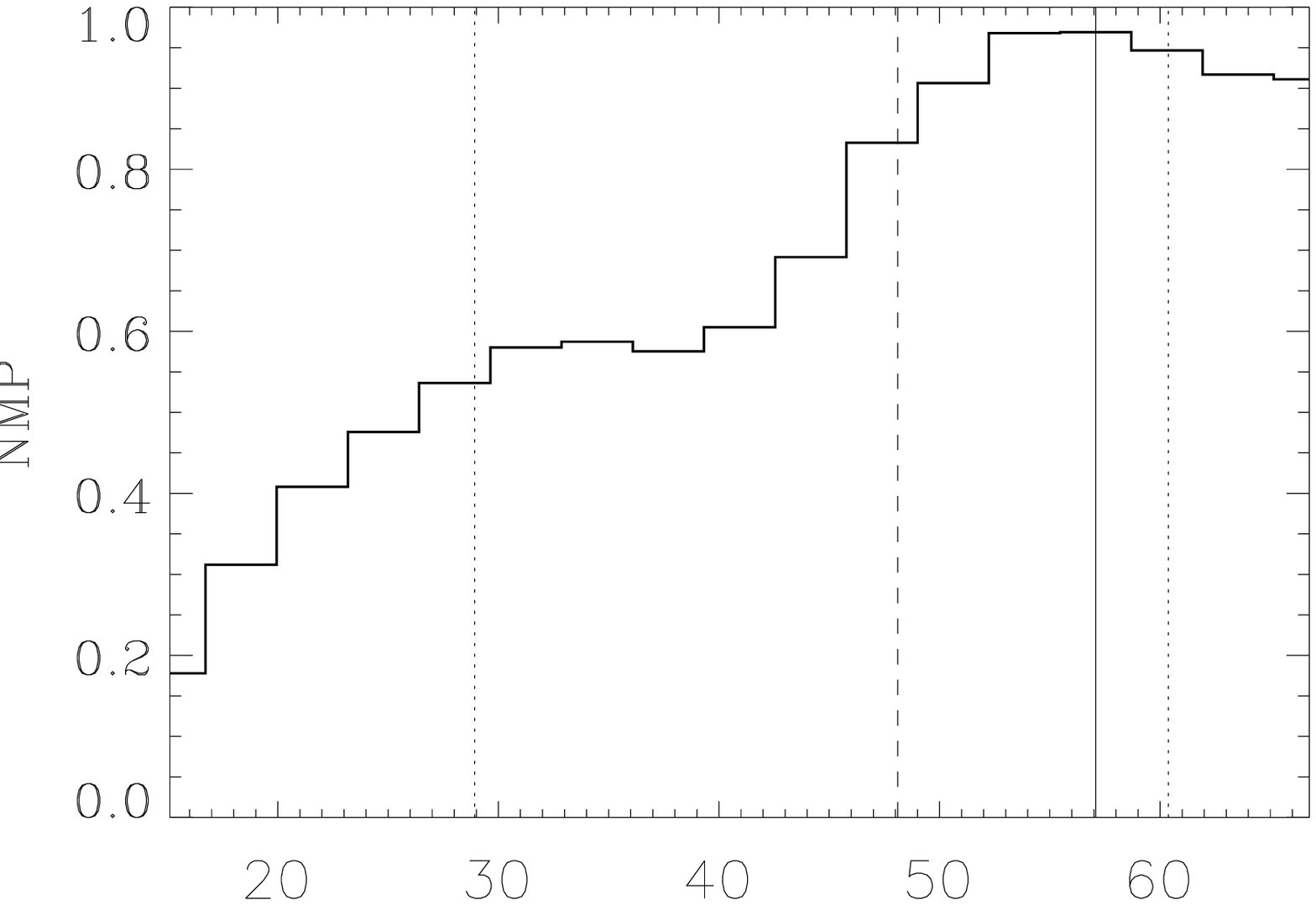}
\includegraphics[width=4cm]{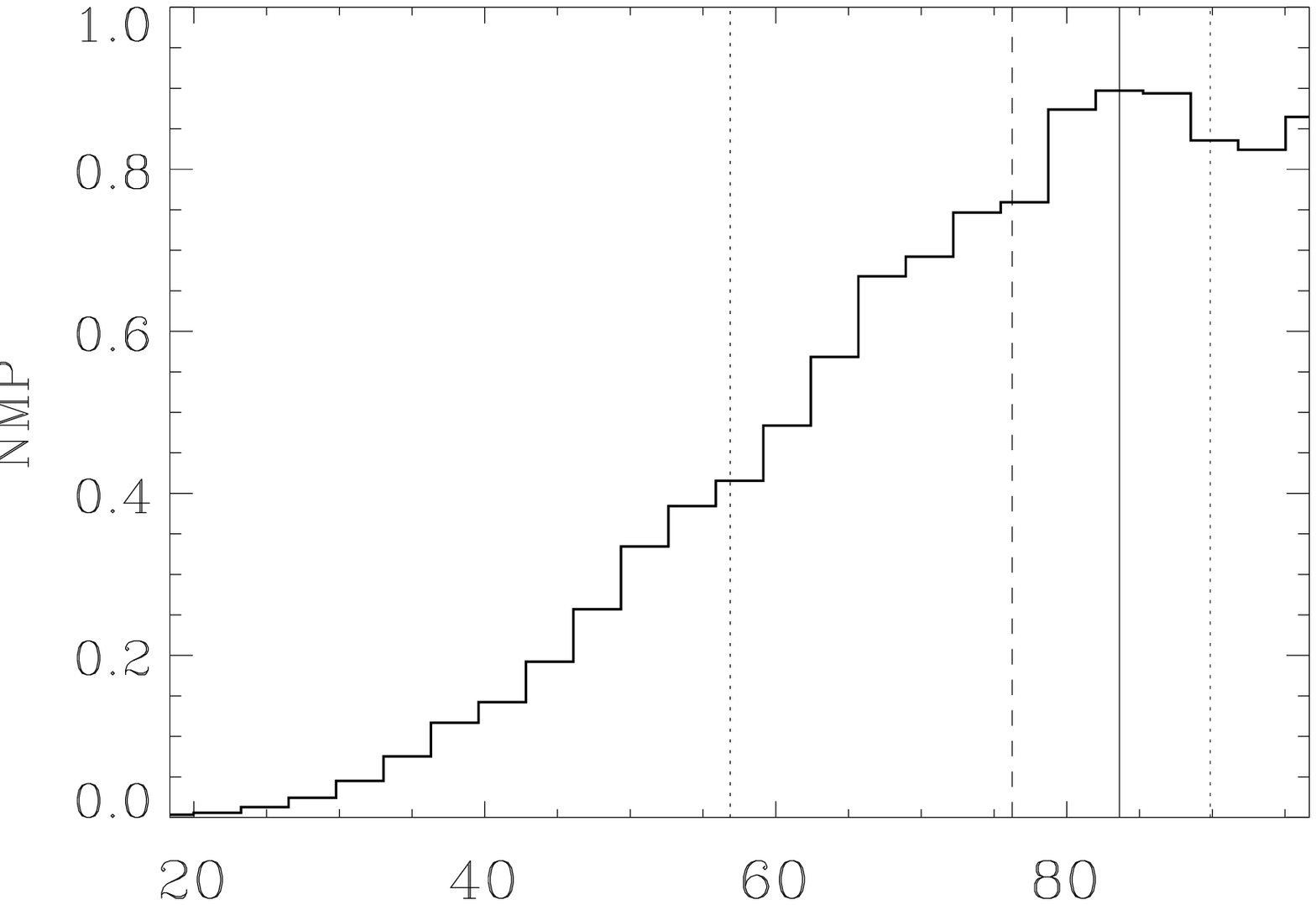}
\includegraphics[width=4cm]{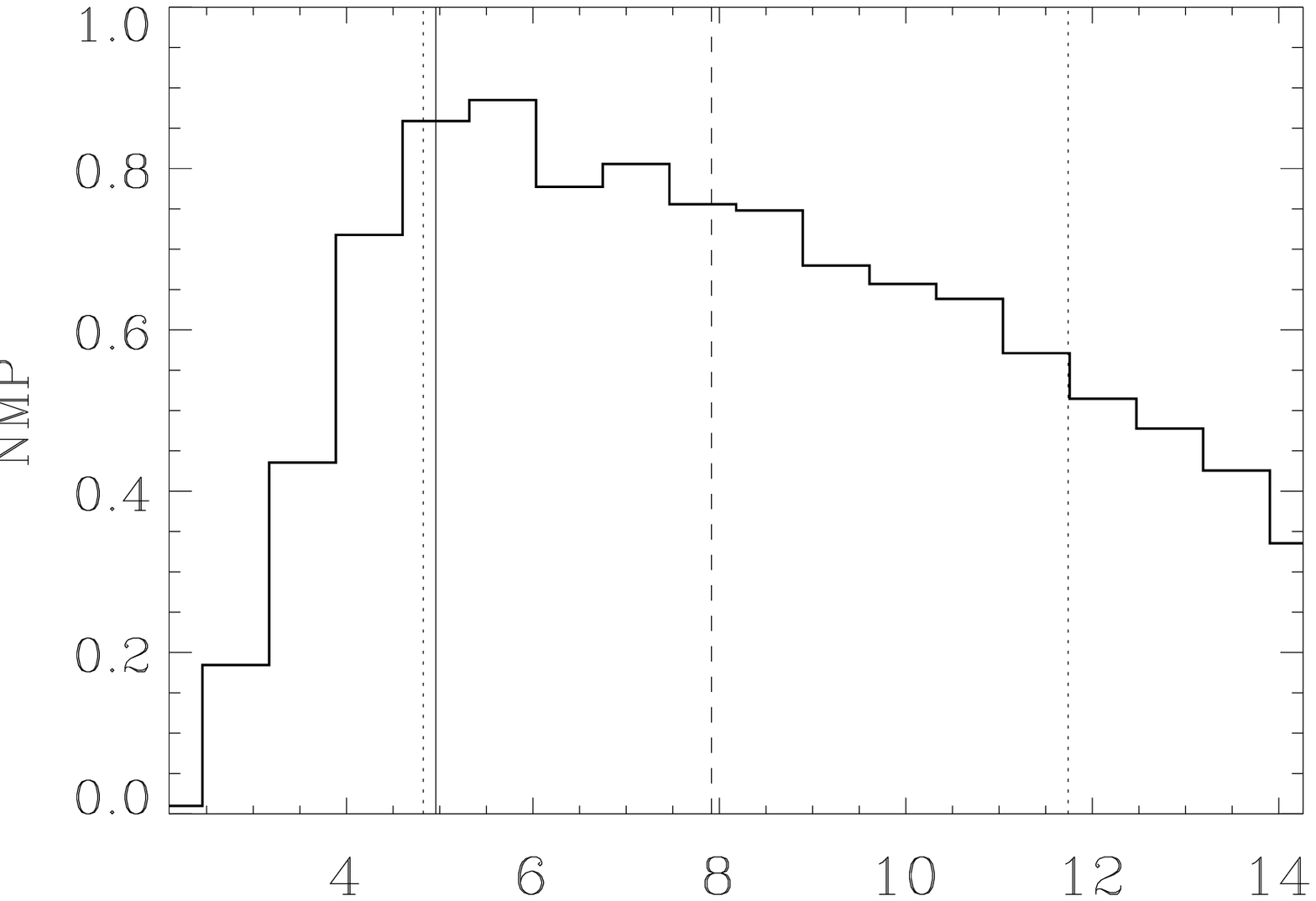}
\includegraphics[width=4cm]{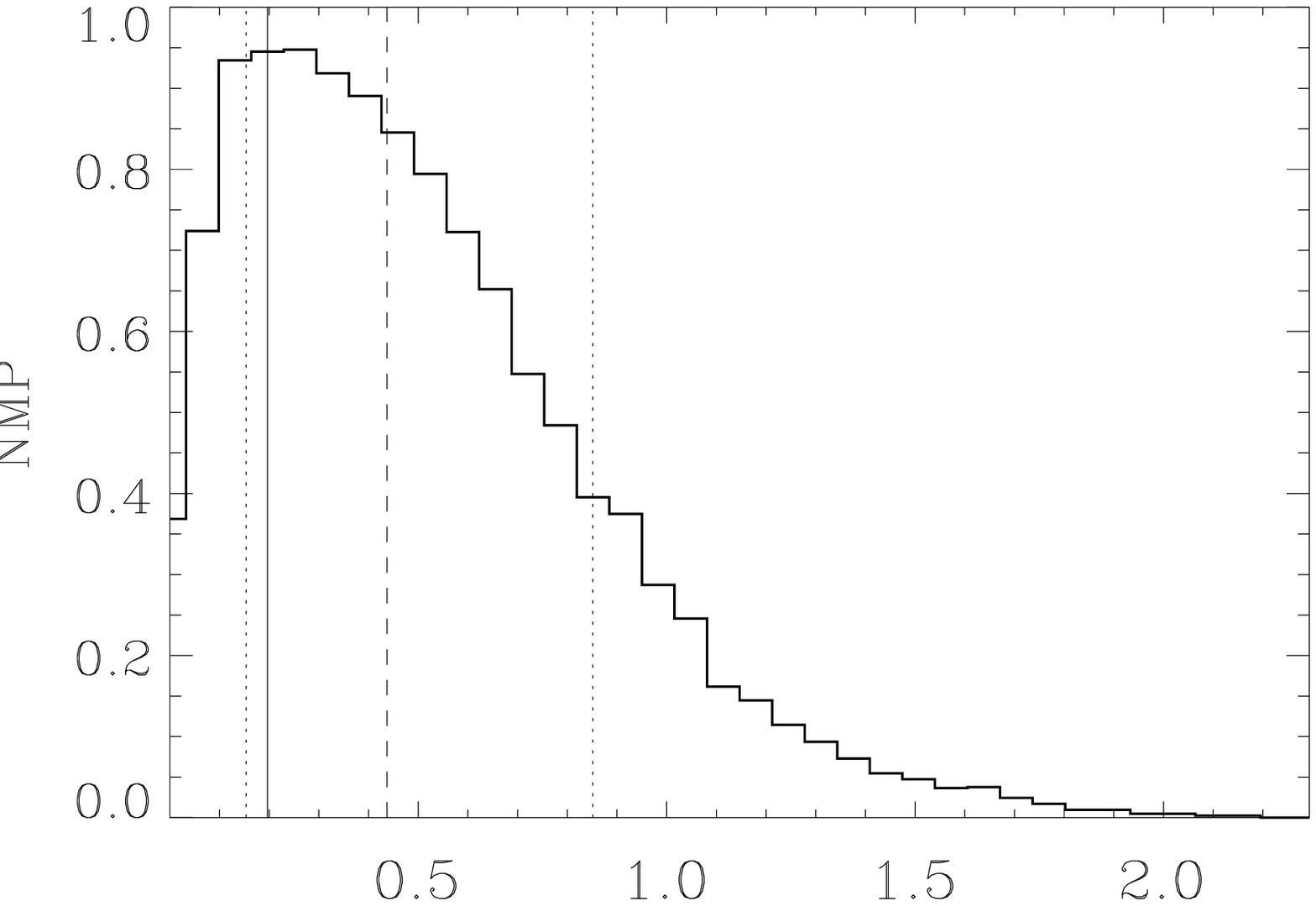}
\includegraphics[width=4cm]{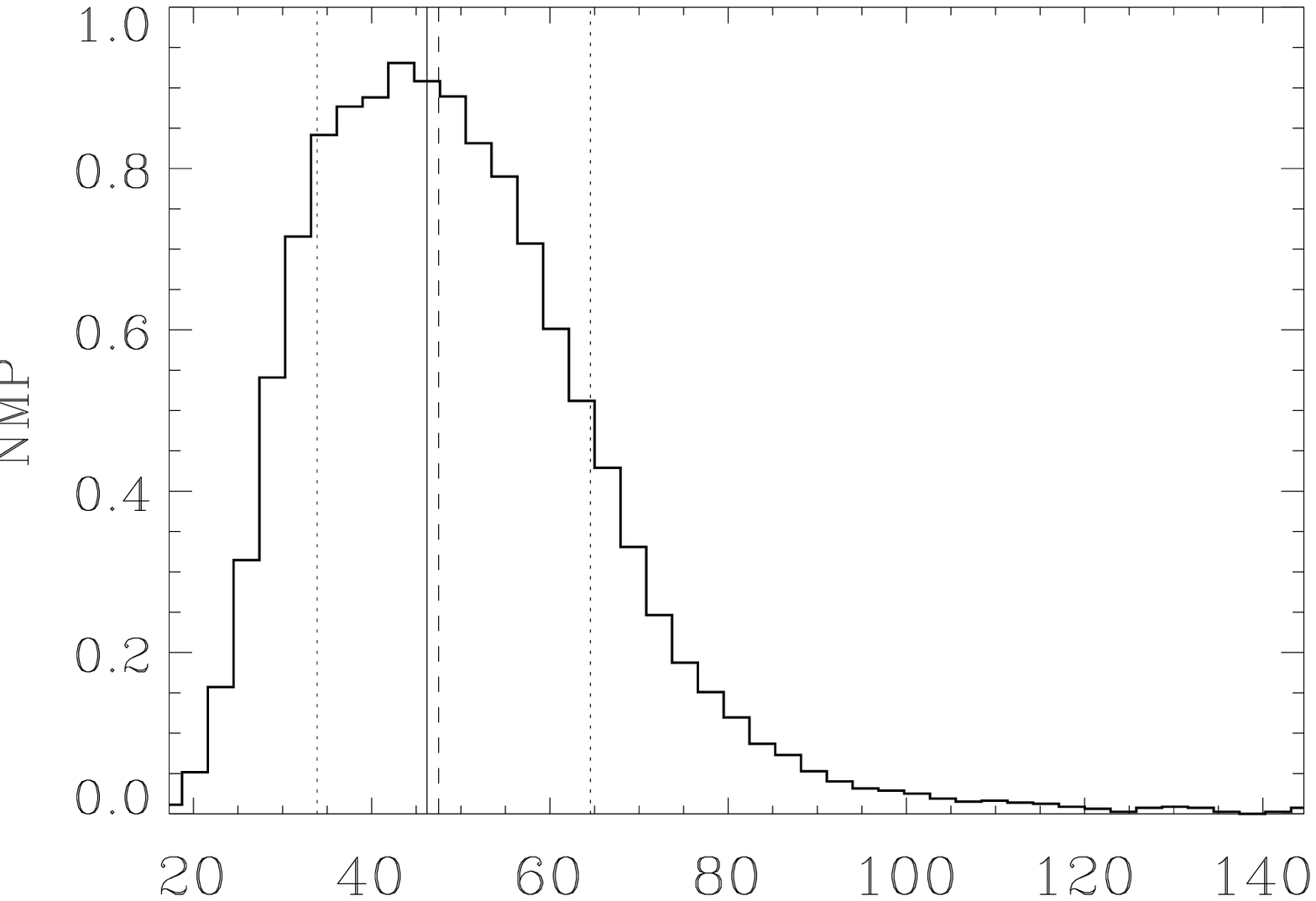}
\includegraphics[width=4cm]{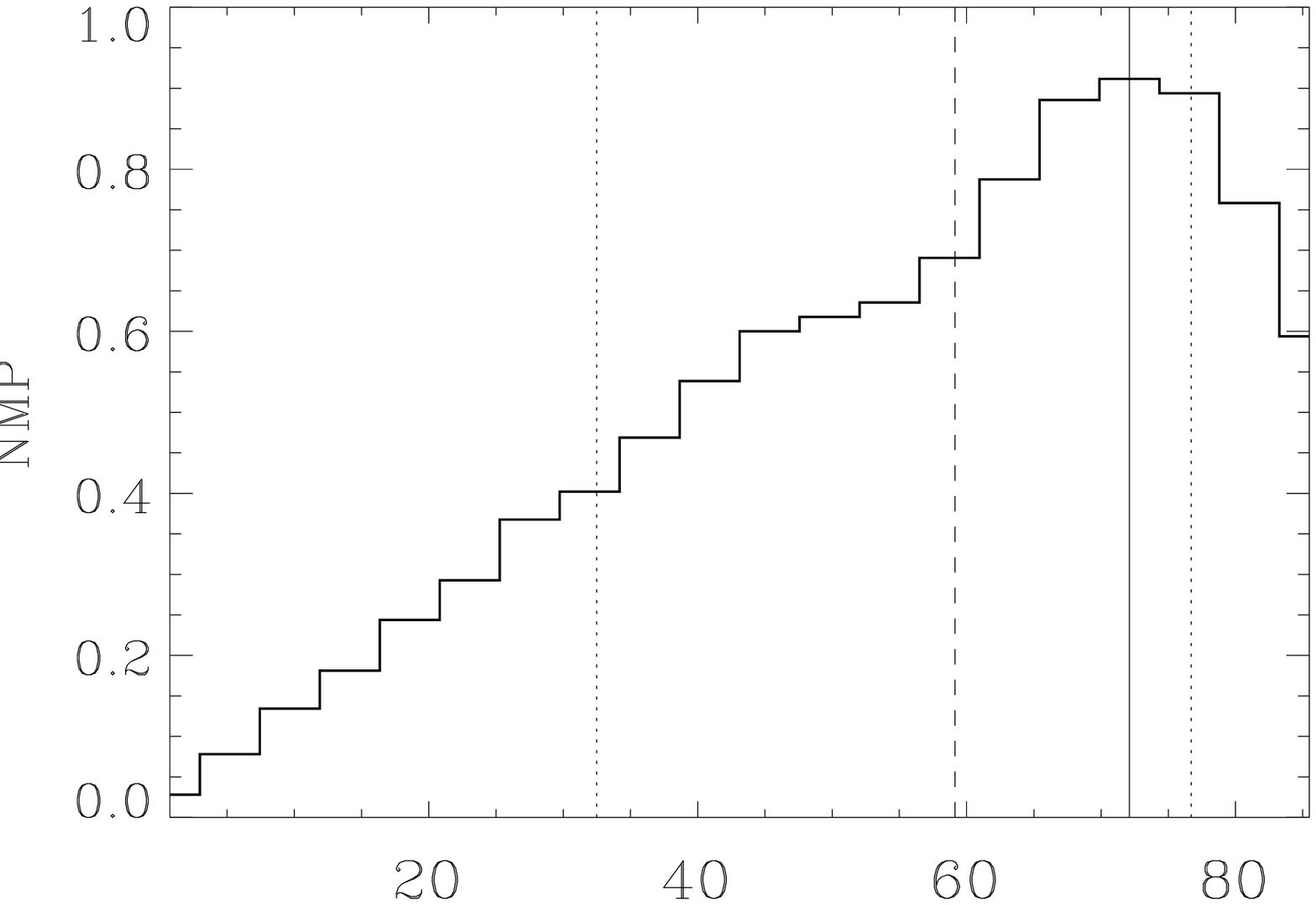}\par} 
\caption{Normalised marginal posteriors (NMP) 
resulting from the fit of NGC 3081. Solid and dashed vertical lines represent the modes and medians of each 
NMP respectively, and dotted vertical lines indicate 
the 68\% confidence level for each parameter around the median.}
\label{ngc3081_distrib}
\end{figure}

The clumpy dusty torus models of \citet{Nenkova08} 
are characterised by six parameters, which are described in Table \ref{parametros}. 
Here we use an interpolated version of the Nenkova models to fit the IR nuclear fluxes reported 
in Table \ref{psf} (considering the SPIRE fluxes as upper limits) using our 
Bayesian inference tool {\it BayesClumpy} \citep{Asensio09} and the uniform priors 
described in Table \ref{parametros}.
The result of the SED fitting are the posterior distributions of
the model parameters (Figure \ref{ngc3081_distrib}), but 
we can translate these results into a single SED. In Figure \ref{ngc3081_fit} we plot 
the model that better fits the IR data, i.e. the maximum-a-posteriori (MAP) model, 
and the model described by the medians of the six posteriors resulting from the fit (see Table \ref{parametros}). 
%See \citet{Ramos09} for a more detailed description on the use of {\it BayesClumpy}.
%The more information provided by the IR SED, the better the posteriors are constrained. 
%The last two columns of Table \ref{parametros} list the medians and modes of the posteriors shown in 
%Figure \ref{ngc3081_distrib}. 
The nuclear NIR, MIR, and FIR emission of NGC 3081, at least up to 160 \micron, is successfully 
reproduced by a relatively broad clumpy torus ($\sigma$=57\degr) with an average number of clouds N$_0$=5 
along the radial equatorial direction, and with a close to edge-on inclination $i$=71\degr.

\begin{table*}
\centering
\begin{tabular}{lclll}
\hline
\hline
Parameter & Abbreviation & Interval & \multicolumn{2}{c}{Fitting results} \\
 & & & Median & Mode \\
\hline
Width of the angular distribution of clouds            & $\sigma$        & [15\degr, 75\degr] &  48\degr$\pm^{12\degr}_{19\degr}$&  57\degr \\
Radial extent of the torus ($R_{o}/R_{d}$)             & $Y$             & [5, 100]           &  76$\pm^{14}_{19}$               &  84      \\
Number of clouds along the radial equatorial direction & $N_0$           & [1, 15]            &  8$\pm^{4}_{3}$                  &  5       \\
Power-law index of the radial density profile          & $q$             & [0, 3]             &  0.4$\pm^{0.4}_{0.3}$            &  0.2     \\
Inclination angle of the torus                         & $i$             & [0\degr, 90\degr]  &  54\degr$\pm^{19\degr}_{25\degr}$&  71\degr \\
Optical depth per single cloud                         & $\tau_{V}$      & [5, 150]           &  59$\pm^{18}_{27}$               &  72      \\ 
\hline
A$_V$ produced by the torus along the line-of-sight (LOS)   & $A_{V}^{LOS}$   & \dots              &  214$\pm^{117}_{85}$ mag         &  162 mag \\
\hline      
\end{tabular}						 
\caption{Clumpy model parameters and $A_{V}^{LOS}$ derived from them. Columns 1 and 2 give the parameter 
description and abbreviation used in the text. Column 3 indicates the 
input ranges considered for the fit (i.e., the uniform priors). Finally, columns 4 and 5 list the 
medians and modes of the posterior distributions shown in Figure \ref{ngc3081_distrib}.}
\label{parametros}
\end{table*}

In Ramos Almeida et al. (2011; hereafter \citealt{Ramos11}) we fitted the NIR/MIR nuclear SED 
of NGC 3081 using the same models as here and obtained $\sigma$, N$_0$, and $i$ values which agree with 
those reported here. On the other hand, the torus radial extent, $Y$, and the index of the clouds radial 
distribution, $q$, change significantly when adding the FIR data. 
The sensitivity of the SED to $q$ for small values of $Y$ is highly reduced. This is because for  
small tori the SED shape does not change noticeably either when the clumps are distributed along the whole extent of the torus ($q$=0)
or highly concentrated in its inner part ($q$=2-3). 
%This lack of sensitivity could result in posteriors that depend
%on the quality of the interpolation technique used by {\it BayesClumpy}. Consequently, whenever the inferred $Y$ is very small, as 
%happened in certain cases when NIR and MIR data only were fitted, the inferred values of $q$ were 
%not completely reliable. However, when including FIR data in the fit of NGC 3081, we obtain a large value of $Y$,
%and then the determination of $q$ is more accurate. 
Here we obtain $q$=0.2, which is characteristic of a flat cloud distribution, and a torus radial extent 
$Y$=84, whereas in \citealt{Ramos11} the resulting values for NGC 3081 were $q$=2.3 and $Y$=22.
However, while in \citealt{Ramos11} we impossed $Y$=[5, 30] as a prior, 
in this work we use $Y$=[5, 100] to take into account cold dust within larger scales. We did the test
of fitting only the NIR and MIR data using the same priors shown in Table \ref{parametros} and obtained
$Y$=35, with the other parameters resulting in similar values as those reported in \citealt{Ramos11}. We also 
performed the fit by considering both the PACS and SPIRE fluxes as upper limits, and the results are
practically the same as those from the NIR/MIR fit. 
Summarising, including the FIR data in the fit of NGC 3081 results in a relatively 
large torus radial extent and flattens the clouds distribution. \\
Using the median value of A$_V^{LOS}$ (in the following we will refer to median values in order to give 
uncertainties at the 68\% confidence level, which are defined around the median) reported in Table 
\ref{parametros} we can derive the column density 
using the Galactic dust-to-gas ratio (N$_H^{LOS}=1.9\times10^{21}~A_V^{LOS}$; \citealt{Bohlin78}). This gives 
N$_H^{LOS}=4.1\pm^{2.2}_{1.6}\times10^{23}~cm^{-2}$, which is compatible with the value derived from 
ASCA X-ray observations of NGC 3081
(N$_H^{X-rays}=6.3\pm0.4\times10^{23}~cm^{-2}$; \citealt{Levenson09}). 
%According to this, 
%the columns of material responsible for the X-ray 
%absorption are slightly larger than those inferred from our IR data, as expected for Seyfert galaxies \citep{Maiolino01}. 
The AGN bolometric luminosity can be obtained from the vertical shift applied to the models to fit the data:
L$_{bol}^{AGN}$ = 2.1$\pm^{1.8}_{0.8}\times$10$^{43}~erg~s^{-1}$ and 
can be directly compared with the 2-10 keV intrinsic luminosity derived 
from the ASCA data, after applying a bolometric correction factor of 20 \citep{Elvis94}:
L$_{bol}^X$ = 1.0$\pm0.2\times$10$^{44}~erg~s^{-1}$. 
The difference between 
L$_{bol}^{AGN}$ and L$_{bol}^X$ is smaller than the one found by \citealt{Ramos11} for NGC 3081. \\
%Thus, L$_{bol}^X$ = 1.0$\pm0.2\times$10$^{44}~erg~s^{-1}$ and  L$_{bol}^X$ $\sim$ 5$\times$L$_{bol}^{AGN}$.
%Combining $L_{bol}^{AGN}$ with the torus bolometric luminosity, which is obtained by integrating the 
%corresponding model torus emission, we derive the reprocessing efficiency of the torus 
%L$_{bol}^{tor}$/$L_{bol}^{AGN}$ = 0.7$\pm^{0.2}_{0.4}$. 
%This means that the torus in NGC 3081 is an efficient reprocessor, 
%since 70\% of the incoming radiation from the AGN is reprocessed in the IR. 
As explained before, the outer size of the toroidal distribution of clouds is defined as $R_{o} = Y R_{d}$, 
where R$_{d}$ scales with L$_{bol}^{AGN}$. Thus, $R_o= 0.4~Y~(L_{bol}^{MIR}/10^{45})^{0.5}$ pc = 4$\pm^{2}_{1}$ pc, which
includes dust at very different temperatures (from T$_d$ to a few Kelvin). This value is larger than the torus 
radius obtained from the fit of NIR and MIR data only (R$_o$=0.7$\pm$0.3 pc; see \citealt{Ramos11}). 
%This value is consistent with 
%interferometric observations of nearby AGN, which indicate that the torus emission 
%only extends out to less than 10 pc (see e.g. \citealt{Jaffe04,Tristram07}).
Finally, we can calculate the torus covering factor, C$_T$. Broader tori 
%is defined as C$_T=1-\int e^{-N_0~e^{(-(i-90)^2/\sigma^2)}}d cos(i)$. Broader tori 
with more clumps will have larger covering factors and viceversa. According to our modelling, 
C$_T$ = 0.8$\pm^{0.5}_{0.3}$, which is among the typical values found for Sy2 galaxies in \citealt{Ramos11} 
and it is compatible with X-ray studies (e.g. \citealt{Ricci11}).

%and means that
%the absorption of the nuclear region in these objects, as derived from optical and IR broad lines, 
%is generally much lower than that obtained from the gaseous column density deduced from the X-rays, if a standard
%Galactic dust-to-gas ratio is assumed. 

\vspace{-0.5cm}

\section{Circumnuclear and extended dust properties}

\begin{figure}
%\centering
\includegraphics[width=9cm]{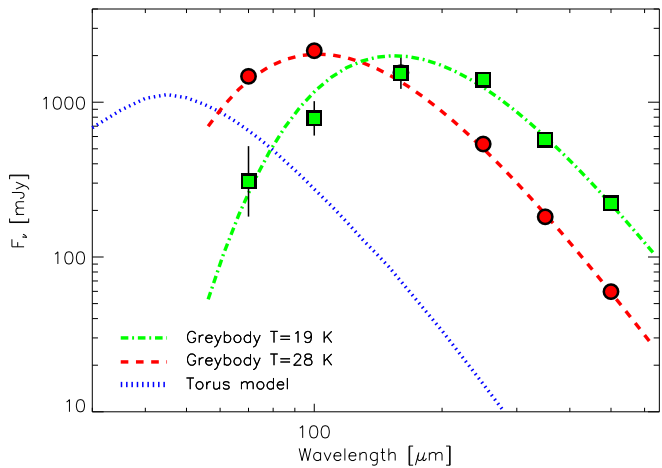}
\caption{Fit of the circumnuclear emission of NGC 3081 with a greybody of T=28$\pm$1 K (red dashed line). 
The green dot-dashed line corresponds to the fit of the disk emission 
with a greybody of T=19$\pm$3 K.}
\label{dust}
\end{figure}

As explained above, the unresolved component of the SPIRE data includes the inner ring r1 apart from the active nucleus
(note the clear bump of emission at $\lambda>$160 \micron~in Figure \ref{ngc3081_fit}).
To characterise the heating source of this component 
we have extracted fluxes in an aperture equals to the maximum size of the PSF in the SPIRE bands (i.e. the FWHM at 500 
\micron; 36.9\arcsec) and then subtracted   
the galaxy background emission measured in an adjacent annulus (first row in Table \ref{aperture}). 
%As explained in Section \ref{intro}, Seyfert galaxies are strong FIR emitters. Depending on the 
%dust temperature, it is possible to distinguish which source is heating this dust. With the previous modelling 
%we have identified the level of IR emission that comes from the 
%torus and is heated by the AGN. 
%In order to investigate if there is any other illuminating source  
%than the AGN in the nuclear region, we have obtained fluxes in an aperture equal to the FWHM of the PSF 
%at 500 \micron~(36.9\arcsec) 
%and then subtracted the contribution of the adjacent galaxy (see Table \ref{aperture}). 
The blue dotted line in Figure \ref{dust} corresponds to the best fitting torus model (MAP), and by subtracting it 
from the latter fluxes, we can isolate the dust emission that is not related to the torus (red dots). 
The best fit that we get for the latter component is a greybody of emissivity $\epsilon$=2 and temperature T=28$\pm$1 K,
which is typical of dust heated by young stars in the galaxy disk. Indeed, based on HST images, \citet{Buta04}
detected $\sim$350 diffuse bright clusters of R$_c\sim$11 pc, with stellar populations younger than 10 Myr. 
In a similar way, we can determine the temperature of the dust in the galaxy disk, by subtracting the fluxes 
obtained in the 36.9\arcsec~aperture from the total galaxy fluxes reported in 
Table \ref{aperture} (green squares). Figure \ref{dust} shows the fit of this component 
with a greybody of $\epsilon$=2 and T=19$\pm$3 K, that is compatible with dust heated by the interstellar radiation field.  

\begin{table}
\begin{tabular}{ccccccc}
\hline
Aperture & F$_{70}$ & F$_{100}$ & F$_{160}$ & F$_{250}$ & F$_{350}$ & F$_{500}$ \\
\hline
\hline
36.9\arcsec & 2124 & 2424 & 1627 & 552  & 187 & 61  \\	
Total flux  & 2432 & 3210 & 3164 & 1950 & 757 & 284 \\
\hline
\end{tabular}
\caption{PACS \& SPIRE flux densities of NGC 3081 in mJy. The total fluxes were obtained
in apertures big enough to include the whole galaxy emission in each band.}
\label{aperture}
\end{table}

We integrated the emission of the two previous components and obtained their IR luminosities 
(1-1000 \micron): L$^{circ}_{IR}$=1.4$\times$10$^9$ L$_{\sun}$
and L$^{disk}_{IR}$=2.2$\times$10$^9$ L$_{\sun}$. Following the equation
M$_d=7.9\times10^{-5}(T/40)^{-6}L_{IR}/L_{\sun}$ (M$_{\sun}$) from \citet{Klaas93}, we can estimate the
dust masses as in \citet{Radovich99}. We obtain a mass of 0.9$\times$10$^6$ M$_{\sun}$ for the circumnuclear 
dust at T=28 K and 1.6$\times$10$^7$ M$_{\sun}$ for the disk. 
%The latter value is smaller than the mass reported by \citet{Radovich99} for 
%the disk of the star-forming Seyfert 2 galaxy NGC 7582 (M$_d$=8$\times$10$^7$ M$_{\sun}$).
The dust mass content of NGC 3081 is $\sim$5 times smaller than the value reported by \citet{Radovich99},  for the disk of
the star-forming Seyfert 2 NGC 7582. 
This mass ratio is twice that of the 
HI masses ($M_{HI}\propto d^2  S_\nu(21cm) \delta v$; \citealt{Giovanelli88}), 
but it is consistent with the relative content of virial mass in both galaxies. 
%The measured HI 21cm line width in the galaxy NGC 3081 is 100 km~s$^{-1}$, which is a factor $\sim$2 smaller 
%that measured in the galaxy NGC 7582. 
%All data were taken from
%the HyperLeda database (http://leda.univ-lyon1.fr/) \citep{Paturel03}."

\vspace{-0.5cm}

\section{Conclusions}

The FIR nuclear luminosity of NGC 3081 (on scales $\leq$1.7 kpc in diameter) can be reproduced 
by warm/cold dust within a clumpy torus heated by the AGN. On larger scales (5.4 kpc), the 
IR emission corresponds to a cold dust component at T=28$\pm$1 K heated by young stars in the galaxy disk, 
likely located in the r1 star-forming ring. On the other hand, the dust located
in the outer parts of the galaxy is heated by the interstellar radiation field (19$\pm$3 K).
These components are coincident with the findings IR studies of nearby Seyfert galaxies  
\citep{Radovich99,Perez01,Bendo10}.

In our previous work using clumpy torus models we fitted 
the NIR-to-MIR SED of NGC 3081, among other Seyfert galaxies, with an interpolated
version of the \citet{Nenkova08} models. In this letter we have repeated the fit after adding the 
nuclear Herschel fluxes to the SED. The FIR data provide information about cooler dust within 
the torus, resulting in a relatively large value of the torus outer radius and a flat radial distribution 
of the clumps.

\vspace{-0.5cm}

\section*{Acknowledgments}

\footnotesize{CRA acknowledges financial support from STFC (ST/G001758/1) and from
the Spanish Ministry of Science and Innovation (MICINN) through project
Consolider-Ingenio 2010 Program grant CSD2006-00070: First Science with the GTC.
AMPG and JIGS acknowledge the Spanish Ministry of Science and Innovation (MICINN) through project AYA2008-06311-C02-01/02.
AAR acknowledges the Spanish Ministry of Science and Innovation through projects AYA2010-18029 
(Solar Magnetism and Astrophysical Spectropolarimetry).
AAH and PE acknowledges support from the Spanish Plan Nacional de Astronom\' ia y Astrof\' isica under grant AYA2009-05705-E.
MP acknowledges Junta de Andalucía and Spanish Ministry of Science and Innovation through projects PO8-TIC-03531 and AYA2010-15169.
PACS has been developed by a consortium of institutes led by MPE (Germany) and including UVIE (Austria); KU Leuven, CSL, IMEC (Belgium); CEA, LAM (France); MPIA (Germany); INAF-IFSI/OAA/OAP/OAT, LENS, SISSA (Italy); IAC (Spain). This development has been supported by the funding agencies BMVIT (Austria), ESA-PRODEX (Belgium), CEA/CNES (France), DLR (Germany), ASI/INAF (Italy), and CICYT/MCYT (Spain). SPIRE has been developed by a consortium of institutes led by Cardiff University (UK) and including Univ. Lethbridge (Canada); NAOC (China); CEA, LAM (France); IFSI, Univ. Padua (Italy); IAC (Spain); Stockholm Observatory (Sweden); Imperial College London, RAL, UCL-MSSL, UKATC, Univ. Sussex (UK); and Caltech, JPL, NHSC, Univ. Colorado (USA). This development has been supported by national funding agencies: CSA (Canada); NAOC (China); CEA, CNES, CNRS (France); ASI (Italy); MCINN (Spain); SNSB (Sweden); STFC (UK); and NASA (USA).
We finally acknowledge the anonymous referee for the useful comments.}

\label{lastpage}

\end{document}